\documentclass[RNAAS]{aastex62}

\begin{document}

\title{Atomic Clocks in Space: A Search for Rubidium and Cesium Masers in M- and L-Dwarfs}

\correspondingauthor{Jeremy Darling}
\email{jeremy.darling@colorado.edu}

\author[0000-0003-2511-2060]{Jeremy Darling}
\affiliation{Center for Astrophysics and Space Astronomy \\
Department of Astrophysical and Planetary Sciences \\
University of Colorado, 389 UCB \\
Boulder, CO 80309-0389, USA}

\begin{abstract}
  I searched for the ground state 6.8 and 9.2 GHz hyperfine transitions of rubidium and cesium toward M- and L-dwarfs that show Rb and Cs optical resonance lines. The optical lines can pump the hyperfine transitions, potentially forming masers. These spin-flip transitions of Rb and Cs are the principal transitions used in atomic clocks (the $^{133}$Cs hyperfine transition defines the second). 
If they are detected in stellar atmospheres, these transitions would provide exceptionally precise clocks that can be used as accelerometers, as exoplanet detectors, as probes of the predictions of general relativity, as probes of light propagation effects, and as a means to do fundamental physics with telescopes.  Observations of 21 M- and L-dwarfs, however, show no evidence for Rb or Cs maser action, and a previous survey of giant stars made no Rb maser detections.  
\end{abstract}


\section{A Rubidium and Cesium Primer} 

Rubidium has atomic number 37 and two common isotopes:  $^{85}$Rb (stable) and $^{87}$Rb (49 Gyr half-life);
the terrestrial isotopic ratio is 72:28 \citep{pringle2017}.  
The $^{87}$Rb  ground state hyperfine transition at 6.83468261090429(9) GHz \citep{bize1999} can form a maser
and is often used as an atomic clock.  
Rb has one valence electron in the $5^2S_{1/2}$ ground state, and the primary 
optical resonance transitions are $5^2S_{1/2}\rightarrow5^2P_{1/2}$ 
and $5^2S_{1/2}\rightarrow5^2P_{3/2}$ at 795 and 780 nm.\footnote{\url{http://steck.us/alkalidata/}}
The 6.8 GHz $^{87}$Rb atomic clock maser relies on the hyperfine structure of the  $^{85}$Rb 
optical resonance lines to selectively filter and optically pump the $^{87}$Rb hyperfine ground states, 
creating a population inversion and promoting maser action \citep{bender1958, davidovits1966}.
The same processes may occur in stellar and sub-stellar
atmospheres, producing a natural 6.8 GHz $^{87}$Rb maser.

The analogous hyperfine cesium transition occurs at exactly 9.192631770 GHz; this
transition defines the second.  Unlike Rb, there is only
one stable isotope of Cs, $^{133}$Cs.  The optical resonance lines at 852.3 and 894.6 nm
correspond to the transitions $6^2S_{1/2}\rightarrow6^2P_{3/2}$ and $6^2S_{1/2}\rightarrow6^2P_{1/2}$.
The pumping of coherent 9.2 GHz Cs emission occurs via collisions with buffer gases of similar pressure
to that found in stellar photospheres, seems to be fairly independent of buffer gas species, and
increases with temperature  \citep{vanier1998}.  Laboratory work was limited by temperature and did not
include ions (although this is not an issue for M- and L-dwarf atmospheres), so the expectation for
stellar Cs maser action is less certain than it is for Rb (but still favorable).

Astrophysical maser action requires the population inversion
of a metastable state, seed photons to amplify (either continuum or spontaneous emission in the maser transition),
and a velocity-coherent amplification pathway.
These processes can obtain in stellar atmospheres, which can be prodigious 
emitters of molecular masers such as SiO, OH, and H$_2$O (typically AGB stars).
%
%
I predict that the conditions in stellar and sub-stellar atmospheres are promising 
for 6.8 GHz $^{87}$Rb and 9.2 GHz $^{133}$Cs maser action.  The \ion{Rb}{1} optical pumping lines 
have been observed in stellar and brown dwarf atmospheres \citep[e.g.,][]{reiners2007}.
Cs is usually detected when Rb is detected, and the Cs maser is collisionally pumped.  

\clearpage

\section{Science with Clocks}

Pulsars have been used as cosmic clocks with great success
\citep[e.g.,][]{backer1986, burke-spolaor2015};
detection and subsequent development of Rb and/or Cs masers would provide
 clocks in new classes of celestial objects.  By tying terrestrial standards to clocks in space, one can 
 test basic physics using telescopes, make (weak) tests of general relativity, detect exoplanets
 via Doppler wobble with unprecedented sensitivity, and,
in concert with {\it Gaia} proper motions, obtain precise three-dimensional kinematics of stars.
%

It is worth stressing that {\it all} spectral lines are clocks.  The power of Rb or Cs hyperfine transitions would lie in their
radio frequency maser action, which enables extremely precise Doppler tracking and astrometry compared to any UV, optical, or IR
transitions (no bright radio emission lines are known in main sequence stars or brown dwarfs).

\section{Astrophysical Rubidium and Cesium}

The optical resonance lines of alkali metals including \ion{Rb}{1} and \ion{Cs}{1} have been detected 
in main sequence stars, brown dwarfs \citep{manjavacas2016}, giant stars \citep[e.g.,][]{garcia-hernandez2006},
and even a candidate Thorne-\.{Z}ytkow object in the Small Magellanic Cloud \citep{levesque2014}.
The $^{85}$Rb and $^{87}$Rb lines are blended and cannot be distinguished in optical spectra, but
the observed presence of other s-process elements, such as Zr, can be used to infer the presence of 
 $^{87}$Rb when the blended \ion{Rb}{1} lines are detected.

Despite the lower abundance of Cs compared to Rb \citep{lodders2003},  Cs absorption lines can be optically thick.
Velocity-coherent column density is key for maser action, and stellar atmospheres satisfy this requirement; the
question for maser production is whether the pumping of $^{87}$Rb or $^{133}$Cs is quenched by collisions.  It is worth noting
that masers have almost always been discovered rather than predicted; they amplify small-scale physical conditions that
may not be representative of the bulk properties of a gas.  Addressing the possibility of Rb or Cs masers in stellar
atmospheres therefore requires observations.  A small Green Bank Telescope survey of giant stars found no
6.8 GHz $^{87}$Rb emission \citep{darling2018}, so I
turn to low-mass stars and brown dwarfs, which provide less distance-dimming and more practical scientific applications
for maser lines, including exoplanet detection and characterization.

\section{Observations}

I selected a sample of 13 M-dwarfs and 8 L-dwarfs where \ion{Rb}{1} and \ion{Cs}{1} are prominent in SDSS DR16 optical spectra \citep{ahumada2020}, 
indicating these elements are abundant and that the maser pumping lines are optically thick.
Using the NSF's Karl G. Jansky Very Large Array\footnote{The National Radio Astronomy Observatory is a facility of the National Science Foundation operated under cooperative agreement by Associated Universities, Inc.} (VLA),
I searched for the 6.8 GHz $^{87}$Rb and 9.2 GHz $^{133}$Cs lines.
VLA observations used the C configuration with integration times of $\sim$10~min, 3~s sampling, and dual circular polarizations.
Bandpasses with 3.91 kHz (0.17  km~s$^{-1}$) channels spanning 8 MHz (351  km~s$^{-1}$) were
centered on the 6.83468261 GHz $^{87}$Rb, and 6.66852 GHz CH$_3$OH transitions, appropriately Doppler shifted to the velocity of each target.
The 9.19263177 GHz Cs observations used 5.208 kHz (0.17 km~s$^{-1}$) channels spanning 16 MHz (522  km~s$^{-1}$).
Synthesized beams ranged from 2.6\arcsec $\times$2.0\arcsec\ to 7.3\arcsec $\times$2.6\arcsec.
I used CASA \citep{CASA} for interferometric flagging, calibration, and imaging.  
Spectral cubes were polarization-averaged and smoothed to 1~km~s$^{-1}$ to achieve 2~mJy rms noise.  No continuum was
subtracted from the cubes.  


\section{Results and Conclusions}

I searched for emission features over a broad velocity range ($\pm 125$~km~s$^{-1}$), taking into account the sometimes high
proper motions of the targets.  No credible maser features were identified.
Table \ref{tab:1} lists the targets, SDSS spectra, and the rms noise of the non-detected transitions.  

These results suggest that Rb and Cs masers are unlikely to occur frequently in M- and L-dwarf atmospheres.
A survey of 10 giant stars and two
globular clusters for the 6.8 GHz $^{87}$Rb maser by \citet{darling2018} likewise made no detections.  I suggest that the search should continue, perhaps
toward other types of stars and in the interstellar medium.

\begin{deluxetable}{lcrcrccc}[t]
\tablecaption{Observations and Results\label{tab:1}}
\tablehead{
  \colhead{Star} & \colhead{SDSS Spectrum\tablenotemark{a}} & \colhead{Coordinates} & \colhead{Spectral} & \colhead{SDSS} & \colhead{$^{87}$Rb} &
     \colhead{$^{133}$Cs} & \colhead{CH$_3$OH}\\[-.07in]
  & & &\colhead{Type} & \colhead{Velocity\tablenotemark{b}}  &   rms\tablenotemark{c} &  rms\tablenotemark{c} & rms\tablenotemark{c}\\[-.07in]
\colhead{}  & & \colhead{(J2000)} & & \colhead{(km s$^{-1}$)} & \colhead{(mJy)} & \colhead{(mJy)} & \colhead{(mJy)} 
}
\startdata
SDSS J000127.84$-$094209.6 & 7167-56604-0196 & 00:01:27.84 $-$09.42.09.73 & M & $-$4.4 & 2.9 & 2.1 & 3.0 \\
2MASS J00191165+0030176 & 4218-55479-0989 & 00:19:11.64 +00.30.17.66 & M & 6.7 & 3.1 & 2.0 & 3.2 \\
SDSS J002226.64+000023.3 & 4219-55480-0199 & 00:22:26.63 +00.00.23.06 & M & $-$9.7 & 2.9 & 2.1 & 3.0 \\
Gaia DR2 2534780227973449728 & 3735-55209-0976 & 01:08:46.43 +00.04.06.94 & M & $-$72.6 & 2.4 & 2.5 & 2.4 \\
SDSS J011453.90+141914.3 & 4664-56192-0948 & 01:14:53.91 +14.19.14.28 & M & $-$32.0 & 2.5 & 2.4 & 2.4 \\
SDSS J012418.33+002242.0 & 4228-55484-0889 & 01:24:18.33 +00.22.42.05 & M & $-$27.0 & 2.4 & 2.6 & 2.4 \\
2MASS J01325911+1312482 & 4666-55832-0755 & 01:32:59.12 +13.12.48.27 & M & $-$3.5 & 2.5 & 2.4 & 2.5 \\
SDSS J015450.56$-$010610.5 & 4233-55449-0242 & 01:54:50.68 $-$01.06.11.01 & M & $-$19.8 & 2.5 & 2.5 & 2.4 \\
SDSS J023100.81+000855.9 & 3647-55945-0818 & 02:31:00.82 +00.08.55.99 & M & $-$45.5 & 2.7 & 2.7 & 2.7 \\
SDSS J023402.03+000623.7 & 3744-55209-0860 & 02:34:02.07 +00.06.22.74 & M & $-$49.3 & 2.4 & 2.5 & 2.3 \\
2MASS J08054990+5113130 & 4528-55559-0368 & 08:05:49.89 +51.13.12.11 & L & $-$15.2 & 2.5 & 2.4 & 2.6 \\
SDSS J081110.35+185527.9 & 4486-55588-0464 & 08:11:10.31 +18.55.27.85 & L & 30.2 & 2.5 & 2.4 & 2.6 \\
2MASS J08175749+1824048 & 4486-55588-0118 & 08:17:57.49 +18.24.04.99 & L & 4.5 & 2.6 & 2.4 & 2.7 \\
SDSS J082906.61+145620.7 & 4503-55563-0828 & 08:29:06.60 +14.56.19.47 & L & $-$5.0 & 2.6 & 2.4 & 2.7 \\
SDSS J083558.28+054830.7 & 4903-55927-0474 & 08:35:58.22 +05.48.30.54 & L & 8.7 & 2.7& 2.4 & 2.8 \\
2MASS J08433323+1024470 & 5284-55866-0967 & 08:43:33.34 +10.24.40.20 & L & $-$11.4 & 2.6 & 2.4 & 2.7 \\
2MASS J10224821+5825453 & 7089-56661-0444 & 10:22:46.83 +58.25.35.19 & L & 15.3 & 2.8 & 2.7 & 2.9 \\
SDSS J103947.32+151251.5 & 5350-56009-0554 & 10:39:47.24 +15.12.51.06 & L & 6.3 & 3.1 & 2.6 & 3.3 \\
SDSS J221451.86+004349.9 & 4200-55499-0873 & 22:14:51.86 +00.43.50.00 & M & $-$73.0 & 2.6 & 2.1 & 2.7 \\
2MASS J22585897+1520461 & 6140-56189-0572 & 22:58:58.87 +15.20.45.06 & M & $-$57.6 & 2.4 & 2.1 & 2.6 \\
2MASS J23522533$-$0944105 & 7166-56602-0319 & 23:52:25.39 $-$09.44.17.52 & M & $-$89.9 & 2.7 & 2.2 & 2.9 \\  
\enddata
\tablenotetext{a}{SDSS plate-MJD-fiber.}
\tablenotetext{b}{Heliocentric optical velocity from SDSS model fit.}
\tablenotetext{c}{Noise per 1.0~km~s$^{-1}$ channel.}
\end{deluxetable}  

\acknowledgments
I thank Z. Berta-Thompson for help with astrometry.

\facilities{VLA}

\software{CASA \citep{CASA}}

\end{document}